\documentclass[apjl]{emulateapj}

\newcommand{\Rs}{R_s}
\newcommand{\Rt}{R_t}

\begin{document}

\title{Dynamical effects of self-generated magnetic fields in cosmic
  ray modified shocks}

\author{D. Caprioli\altaffilmark{1}, P. Blasi\altaffilmark{2,3},
  E. Amato\altaffilmark{2}, M. Vietri\altaffilmark{1}}
  
\altaffiltext{1}{Scuola Normale Superiore, Pisa, Italy}
\altaffiltext{2}{INAF/Osservatorio Astrofisico di Arcetri, Firenze,
  Italy}
\altaffiltext{3}{Fermilab, Center for Particle Astrophysics, USA }


\begin{abstract}
Recent observations of greatly amplified magnetic fields ($\delta B/B
\sim 100$) around supernova shocks are consistent with the
predictions of the non-linear theory of particle acceleration (NLT),
if the field is generated upstream of the shock by cosmic ray induced
streaming instability. The high acceleration efficiencies and large
shock modifications predicted by NLT need however to be mitigated to
confront observations, and this is usually assumed to be accomplished
by some form of turbulent heating. We show here that magnetic fields
with the strength inferred from observations have an important
dynamical role on the 
shock, and imply a shock modification substantially reduced with respect
to the naive unmagnetized case. The effect appears as soon as the
pressure in the turbulent magnetic field becomes comparable with the
pressure of the thermal gas. The
relative importance of this unavoidable effect and of the poorly known
turbulent heating is assessed. More specifically we conclude that even
in the cases in which turbulent heating may be of some importance, the
dynamical reaction of the field cannot be neglected, as instead is
usually done in most current calculations. 
\end{abstract}


\maketitle

\section{Introduction}

The supernova remnant (SNR) paradigm for the origin of galactic cosmic
rays is based on the assumption that at least $\sim 10-20\%$ of the
kinetic energy of the expanding shell is converted into cosmic rays
(CRs). Moreover, as recent observations have proved, the magnetic
field (MF) in the shock vicinity is amplified by a large amount
(e.g. \cite{ballet}) as would be expected if cosmic rays induce
streaming instability (SI) upstream of the shock. We stress that such
MF amplification is required to accelerate protons up to
$\sim 10^6$ GeV. The need for a satisfactory and self-consistent
description of these points is sufficient to justify the
development of a NLT of particle acceleration. 

The developments of the theory are summarized in (\cite{druryrev},
\cite{je91}, \cite{maldrury}). The kinetic theory for arbitrary
diffusion coefficients (\cite{amato1}), and even in the case of
self-generated MFs (\cite{amato2} and \cite{elli06}) has 
been recently developed.

All approaches to NL shock acceleration find that the large
pressure of the accelerated particles decelerates the incoming gas, and
leads to total compression factors that scale with the Mach number of
the shock as $\Rt\sim M_0^{3/4} \sim 20-50$. Such large shock
modifications however are at odds with observations which are better 
fit by $\Rt\sim 7-10$. The problem with larger values of $\Rt$
resides in both the estimated distance between the forward and reverse
shocks (\cite{warren}) and in the fit to multifreqeuncy
observations with concave spectra (VBK05 and references therein). The
reduction in the compression factor is almost invariably attributed to
turbulent heating (TH) in the precursor (\cite{bv97} and
later) as due to damping of waves on the background plasma
(\cite{mck-v82}, hereafter MKV82). 
In fact this mechanism was originally proposed in
order to keep the MF amplification in the linear regime
(e.g. $\delta B/B\ll 1$), but is now commonly applied to cases in
which $\delta B/B\gg 1$.  
Unfortunately, the heating process is quite model dependent and
even its applicability to situations of interest for SNRs can and
should be seriously questioned. The effectiveness of the heating
process can easily be reduced to negligible levels or artificially
amplified to unphysical levels.   

As mentioned above, a breakthrough in the field has recently been
provided by X-ray observations: the detection of X-ray bright
filaments in the outskirts of some SNRs allows one to infer the
strength of the
MF in these filaments, found to be $B\sim 100-500\mu
G$. Such strong fields are generally attributed to the SI induced by
CRs efficiently accelerated at the shock front, although alternative
explanations have been proposed (\cite{gj07}). In Tab. \ref{tab} we
list some SNRs with estimated MFs: $u_0$ is the shock
velocity, $B_2$ is the MF downstream of the shock as
inferred from the X-ray brightness profile and
$P_{w2}=B_2^2/(8\pi\rho_0 u_0^2)$ is the downstream magnetic pressure
normalized to the bulk one. The values of the parameters are from
\cite{P+06} and from VBK05 (in parenthesis in Tab. \ref{tab}).

\begin{table}
\begin{center}
\caption{\label{tab}Parameters for 5 well known SNRs.}
\begin{tabular}{cccc}
\tableline
SNR & $u_0 (km/s)$ & $B_2(\mu G)$ & $P_{w2}\times 10^3$\\
\tableline
Cas A& 5200 (2500)& 250--390 & 32 (36) \\
Kepler & 5400 (4500)& 210--340 &  23 (25)\\
Tycho & 4600 (3100)& 300--530 &  27 (31)\\
SN 1006 & 2900 (3200)& 91--110 & 40 (42) \\
RCW 86 & (800) & 75--145 & 14-35 (16-42)\\
\tableline
\end{tabular}
\tablecomments{For SNRs discussed by \cite{P+06} we used 
$\rho_0=0.1\,m_p/cm^3$ in the case of SN 1006  and $\rho_0=0.5\,m_p/cm^3$ 
in the other cases, while VBK05 provide directly $P_{w2}$.}
\end{center}
\end{table}
We show below that for the field stregths inferred for SNRs, the
magnetic pressure is comparable or even in excess of the thermal
pressure of the background plasma and that whenever this happens the
dynamical reaction of the field on the fluid is such that the
compression factors are substantially reduced and fall in the range
suggested by observations. It is crucial to keep in mind that, 
contrary to the TH, which can be either suppressed or amplified 
by changing parameters on which there is little or no control,
the feedback of the self-generated turbulent MF on the plasma is not
model dependent and must be included. 

\section{Dynamics of a magnetized CR modified shock} 
The reaction of accelerated particles upstream of the shock leads to
the formation of a precursor, in which the fluid speed 
decreases while approaching the shock. One can describe this effect by
introducing two compression factors $\Rt=u_0/u_2$ and
$\Rs=u_1/u_2$, where $u$ is the fluid speed and the indexes '$0$',
'$1$' and '$2$' refer to quantities at upstream infinity, upstream and
downstream of the subshock respectively. 

The most general equations of conservation of mass, momentum and
energy in the stationary case for a parallel shock are: 
\begin{equation}
\frac{\partial}{\partial x}\left( \rho u \right) = 0,\,\,\,\,\,\,\,
\frac{\partial}{\partial x} \left( \rho u^2 + p + p_c + p_w \right)=0, 
\label{eq:massmomentum}
\end{equation}
\begin{equation}
\frac{\partial}{\partial x} \left( \frac{1}{2}\rho u^3 + 
\frac{\gamma p u}{\gamma-1} + F_w \right) = -u \frac{\partial
  p_c}{\partial x}.  
\label{eq:energy}
\end{equation}

As usual, $\rho$, $u$, $p$ and $\gamma$ stand for density, velocity,
pressure and the ratio of specific heats of the gas, while $p_w$ and
$F_w$ represent the pressure and energy flux in the form of Alfv\'en
waves. $p_c$ is the CR pressure. The continuity of the distribution
function of the accelerated particles through the subshock implies
that the CR pressure is also continuous ($p_{c1}=p_{c2}$), and that
the term $\partial p_c/\partial x$ gives null contribution when
Eq.~\ref{eq:energy} is integrated from $x=0^-$ to $x=0^+$. All
previous equations at the subshock read as the usual Rankine-Hugoniot
relations at a magnetized gaseous shock. 
  
In order to treat the presence of Alfv\'en waves correctly, we use the
approach of \cite{vs99} (hereafter VS99), considering two upstream
wave trains with helicities $H_c=\pm 1$, and their respective
downstream counterparts. If $\delta \vec B_{i}$ is a mode of the
MF perturbation, we write the velocity perturbation as
$\delta \vec u=-H_c\frac{\delta\vec B}{\sqrt{4\pi\rho}}$. Neglecting
the electric field contribution, which is of order $(u/c)^2$, the
magnetic pressure and the energy flux, which for Alfv\'en waves is the
sum of the normal component of Poynting vector $\vec u\times\delta\vec
B\times\delta\vec B/4\pi$ and the transverse kinetic energy flux
$\rho\delta\vec u^2/2$, are
\begin{equation}\label{Fw}
p_w=\frac{1}{8\pi}\left(\sum_i\delta\vec B_i\right)^2;\quad
F_w=\sum_i\frac{\delta \vec B_i^2}{4\pi}\tilde{u_i}+
p_w u\,,
\end{equation}
having posed $\tilde{u_i}=u+H_{c,i} v_A$. 
The upstream magnetic turbulence typically shows two opposite
helicities (\cite{bl2001}), each of which splits into a reflected and a
transmitted wave crossing the subshock. 
According to VS99, the transmission and reflection
coefficients, in the limit $M_A^2\gg \Rs$ (large Alfv\'enic Mach
number), do not depend on $H_c$ and read 
\begin{equation}
\label{TR}
T \simeq(\Rs+\sqrt{\Rs})/2\,; \qquad 
R \simeq(\Rs-\sqrt{\Rs})/2\,. 
\end{equation}
For a typical supernova shock, the Alfv\'enic Mach number is
$M_{A,1}=u_1/v_A \geq 100$, hence in the following we adopt these
coefficients and neglect $v_A$ with respect to the fluid velocity in
Eq.~\ref{Fw}. For each $H_c$ we therefore have $\delta B_2/\delta
B_1=T+R=\Rs$ and thus $p_{w2}=p_{w1} \Rs^2$.  

As pointed out above, the subshock can be viewed as a simple shock in
a magnetized gas, therefore the pressure jump is (VS99) 
\begin{equation}\label{p2p1}
\frac{p_2}{p_1}=\frac{(\gamma+1)\Rs-(\gamma-1)+(\gamma-1)(\Rs-1)\Delta}
{\gamma+1-(\gamma-1)\Rs}\;, 
\end{equation}
with $\Delta$ defined as:  
\begin{equation}
\Delta=\frac{\Rs+1}{\Rs-1}\frac{\left[p_{w}\right]}{p_1}-
\frac{2\Rs}{\Rs-1}\frac{\left[F_w\right]}{p_1 u_1}\;. 
\end{equation}
Using the expressions for $T$ and $R$ (Eq.~\ref{TR}) we get
\begin{equation}\label{delta}
\Delta=(\Rs-1)^2\frac{p_{w1}}{p_1}+\Rs~\frac{\vec B_-\cdot\vec
B_+}{2\pi p_1}\,. 
\end{equation}
Following VS99, we assume that the two opposite--propagating waves
carry MFs $\vec B_\pm$ displaced in such a way that $\vec
B_-\cdot\vec B_+=0$. This is not the most general configuration, but
it is nevertheless rather common since it occurs when: 1) there is
only one wave train, 2) when the two fields are orthogonal, and 3) on
average, when the relative phase between the wave trains is arbitrary.  

At this point we normalize all quantities to the ones at upstream
infinity: $U(x)= u(x)/u_0$, $P_w(x) = p_w/\rho_0u_0^2$ and
$P(x)=p(x)/\rho_0u_0^2=\frac{U(x)^{-\gamma}}{\gamma M_0^2}$. In the
latter, we used the assumption of adiabatic heating in the precursor
and the conservation of mass. 

Substituting Eq.~\ref{p2p1}, Eq.~\ref{delta} and the above expression
for $P(x)$ in the equation for momentum conservation, the compression
factors $\Rs$ and $\Rt$ are related through the equation 
\begin{equation}\label{rsrt}
\Rt^{\gamma+1}=\frac{M_0^2\Rs^\gamma}{2}\left[\frac{\gamma+1-\Rs(\gamma-1)}  
{1+\Lambda_B}\right],
\end{equation}
which is the same as the standard relation apart from the factor
$(1+\Lambda_B)$, with  
\begin{equation}
\Lambda_B=W\left[1+\Rs\left(2/\gamma-1\right)\right]\;,
\end{equation}
and $W=P_{w1}/P_1$. It is clear that the net effect of the magnetic
turbulence is to make the fluid less compressible, noticeably reducing
$\Rt$ if $W=P_{w1}/P_1$ is of order 1. 
Moreover, the pressure and temperature jumps at the subshock are
enhanced (Eq.~\ref{p2p1}). 

We should notice that if one naively assumed that $F_w=3up_w$
everywhere, neglecting the $T$ and $R$ coefficients needed to satisfy
Maxwell equations at the subshock, one would obtain
$\Delta'=[(\Rs-1)^2-2\Rs] W < \Delta\,,$ 
leading to an incorrect pressure jump. This approach, adopted by
\cite{elli06}, also leads to a less marked decrease of $\Rt$, since
$\Lambda_B'=W\left[1+\Rs\left(3/\gamma-2\right)\right]$.

\section{Confronting observations}

Here we show that the magnetization levels estimated in SNRs as
reported in Tab. \ref{tab} imply that $W\geq 1$, so that the dynamical
feedback of the amplified MF needs to be taken into
account. 

\begin{figure}
\includegraphics[width=.45\textwidth]{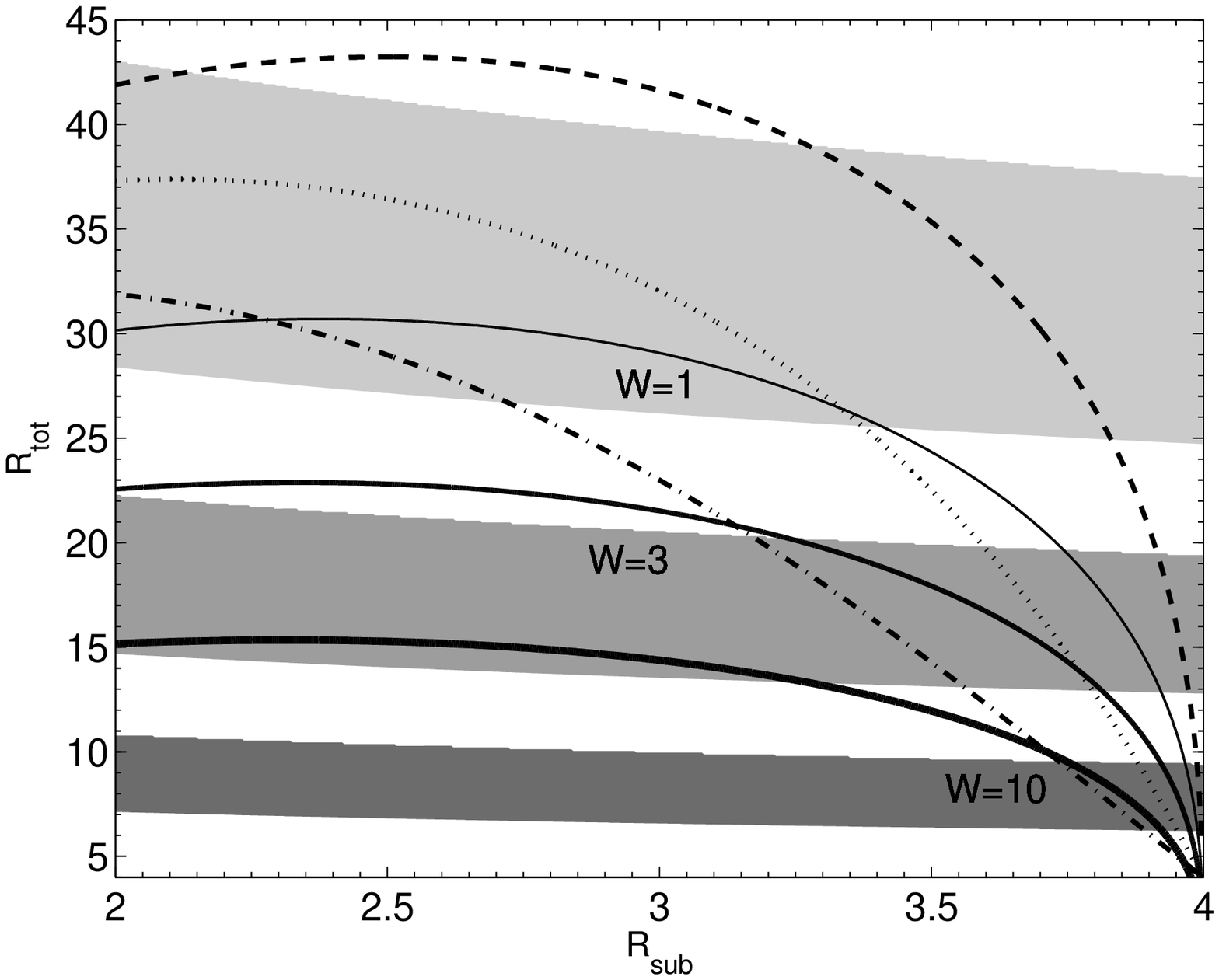}
\caption{$\Rs$-$\Rt$ for $W=1,3,10$ (intersections of
  solid lines with corresponding shadowed regions) and for 
  $\chi=200,500$ (dash-dotted and dotted lines). The dashed line
  represents the case $W=0$ (see text for details).} 
\label{fig:M100}
\end{figure}

In Fig. \ref{fig:M100} we plot $\Rt$ versus $\Rs$ for $M_0=100$: the
three shadowed regions  represent the relation between $\Rs$ and
$\Rt$ for fixed $P_{w2}\in [0.02,0.04]$ and $W=1,~3,~10$:
$P_{w2}/\Rs^2 = P_{w1} = W P_{g1}
=W\left(\Rt/\Rs\right)^\gamma/(\gamma M_0^2)\,$.  

The three solid lines represent the relation $\Rt-\Rs$ for the three
given values of $W$ as given by Eq. \ref{rsrt}; the dashed line refers
to $W=0$, when $p_w$ is not included.    

The compression factor lies at the intersection between the curve and
the shadowed region for a given value of $W$. If $W<0.7$ there are no
intersections. This implies that the values of the magnetic
pressure inferred from observations require substantial MF
amplification upstream, 
and that the conservation equations are affected by the dynamical
reaction of the field. Only values $W\geq 3$ are compatible with the
whole range $0.02\leq P_{w2}\leq 0.04$ inferred from observations. 
This means that in order to account for the inferred values of
$B_2$, the magnetic pressure must be \emph{at least} comparable to the
gas pressure, and thus its dynamical role cannot be neglected. From
Fig. \ref{fig:M100} one also sees that the magnetic reaction leads to
values of $\Rt$ lower by roughly a factor $\sim 2$ compared with the
case $W=0$. We will comment further on this point below. 

Up until this point we never used the physically crucial point that
the observed fields may be generated through a cosmic ray
induced SI upstream of the shock. The instability may operate in the
resonant (\cite{bell78}) and in the non-resonant (\cite{bell2004})
regime.  

The growth rates of these different modes can be easily estimated
only in the context of quasi-linear theory. Given the difficulty in
deriving this information in the general non linear case, here we
assume the following general relation between the pressures of CRs and
wave pressure upstream of the subshock: 
\begin{equation}\label{defchi}
\xi_1 = \chi P_{w1},
\end{equation}
where $\xi(x) = p_{cr}(x)/\rho_0u_0^2$ is the normalized CR
pressure. For resonant SI, one has $\chi\simeq M_A=u_0/v_A$, while for
non-resonant modes, $\chi\sim 4c/u_0$. In both cases, for typical
values of the parameters, one obtains $200\leq\chi\leq 500$.  

From Eq. \ref{eq:massmomentum} applied to the
precursor, namely between upstream of the subshock ($x=0^-$) and
upstream infinity, we can write:  
\begin{equation}
\frac{\Rs}{\Rt} + \frac{1}{\gamma M_0^2} \left[ \left(
  \frac{\Rt}{\Rs}\right)^\gamma -1 \right] + P_{w1}(1+\chi) = 1.  
\label{eq:prec}
\end{equation}
The physical values of $\Rs$ and $\Rt$ for a given $\chi$ are obtained
by determining the intersection of the corresponding curve with that
obtained for a given value of $W$ at the subshock. Whether the
solution reproduces the estimated value of $P_{w2}$ depends on whether
the intersection falls within or outside the shadowed region for
the same $W$ in Fig. \ref{fig:M100}. The dash-dotted and the dotted
line show the results for $\chi=200$ and $500$ respectively: it is
evident that the chosen values of $\chi$ allow for a consistent
explanation of the downstream magnetic pressures as inferred from
observations, and, 
equally important, lead to compression factors which are much lower
than those predicted by the standard NLT (\cite{bv97} and papers that
followed).   

\section{Heating in the precursor}

The strong shock modification predicted by NLT when the magnetic
pressure is ignored is usually assumed to be somewhat mitigated by
heating of the precursor as a result of damping of Alfven waves
(\cite{v-mck81},hereafter VMK81, and MKV82) on the background gas. 
Other phenomena (for instance acoustic instability) may also lead to heating of the
precursor. In the original description, that remained basically
unchanged to the present time, VMK81 assumed that the rate of
damping ($\Gamma$) equals the rate of growth ($\sigma$) of Alfven
waves. The main implication of this assumption is that the growth of
the waves never reaches the non-linear regime, which is in fact the very
reason why the mechanism was invoked in the 80's. The recent observations 
prove that waves can grow to $\delta B/B\gg 1$. It is therefore at
least not self-consistent to apply the standard treatment for TH to
situations in which MF 
amplification to the non-linear regime takes place. In a minimal attempt
to include faster growth one may assume that $\Gamma=\alpha \sigma$,
with $\alpha<1$. Following MKV82 and \cite{be99} one can then 
obtain a generalized relation between $\Rt$ and $\Rs$ in the form
\begin{equation}\label{rsrtTH}
\Rt^{\gamma+1}=\frac{M_0^2\Rs^\gamma}{2}\left[\frac{\gamma+1-\Rs(\gamma-1)}  
{(1+\Lambda_B)(1+\Lambda_{TH})}\right],
\end{equation}
where
\begin{equation}	
\Lambda_{TH}=\alpha (\gamma-1)\frac{M_0^2}{M_A}
\left[1-\left(\frac{\Rs}{\Rt}\right)^\gamma\right]\,,
\end{equation}
which becomes equivalent to the standard Eq. 50 of \cite{be99} only for
$\alpha=1$. Now it is easy to check that for typical values of $\Rs$
and $\Rt$ $\Lambda_{TH}>\Lambda_B$ if $\alpha\gtrsim
3W\frac{M_A}{M_0^2}$. For instance for $M_A\sim 10^3$ and $M_0\sim
100$ one requires $\alpha$ to be of order unity. In this case
however it is not easy to amplify the MF to $\delta B\gg
B_0$. If $\alpha$ is appreciably smaller than unity, the main process
for the smoothening of the precursor is the dynamical reaction of the
self-generated MF. In both cases the role of TH can be
seriously questioned. 

A deeper look at the physical processes that may result in the heating
of the precursor make the role of TH even more uncertain: in the
original papers of VMK81 and MKV82 the Alfven heating was
considered as a result of non-linear Landau damping in a gas in the
hot coronal phase. The authors reached the conclusion that the damping
is important if $u_0\ll 4000 km/s (T_0/5\times 10^5 K)^{1/2}$, where
$u_0$ is the shock velocity and $T_0$ is the temperature of the
unshocked gas. It is all but clear whether for the velocities and
temperatures that apply to the SNRs in Tab. \ref{tab}, non-linear
Landau damping is such to lead to $\alpha\sim 1$. We stress that at
the same time, $\alpha$ cannot be too close to unity, otherwise TH
inhibits the growth of $\delta B$ to the observed levels. 

Other types of turbulent heating may be at work but a quantitative
analysis of these phenomena is lacking at the present time. The
expression for $\Lambda_{TH}$ is however rather general, in that we
did not specify the mathematical form of the growth and damping
rates. Therefore we expect to draw similar conclusions in terms of the
parameter $\alpha$. 

This section strongly suggests that, contrary to the common wisdom,
the most likely reason for the smoothening of the precursor is the
dynamical reaction of the generated MFs rather than some
form of non adiabatic heating in the precursor.

\section{Conclusions}
It is well known that the effect of a MF is in general
that of reducing the plasma compressibility. We showed here that when
applied to a cosmic ray modified shock, 1) this finding implies that 
CR induced SI is adequate to explain the magnetization inferred from 
X-ray observations; 2) the downstream
MFs imply that $W\sim 1-10$, so that the field becomes  
dynamically important, since this happens whenever the magnetic
pressure upstream becomes comparable with the gas pressure, namely
when $W>0.7-1$; 3) the dynamical reaction of the
MF reduces the compression in the precursor, leading to
smaller (larger) values of $\Rt$ ($\Rs$) in agreement with the values
required to explain the distance between forward and reverse shock and
the multifrequency observations of several SNRs; 4) this effect comes
from first principles, though in our calculations we restricted our
attention to the case of Alfven waves, and is not affected by the huge
uncertainties typical of TH; 5) an efficient TH may smoothen the
precursor if $\alpha$ is close to unity, but in this case it is likely
to inhibit the growth of the field to $\delta B\gg B_0$. 

Although the underlying physics is well known, the dynamical effect of
the magnetic pressure has not been included in the calculations of 
multifrequency emission from SNRs (\cite{bv97} and successive papers),
so that the strong modifications predicted by NLT had to be
compensated by assuming TH. The only exception that we are aware of is
the recent work by \cite{elli06}, in which the authors perform Monte Carlo
simulations of the particle acceleration process including the
pressure of self-generated MFs. In such simulations, which
represent the state of the art in the field, however, thermal
and accelerated particles are treated in the same way, therefore the
condition $W\sim 1$ could not be tested. We suspect that for this
reason the smoothening of the precursor was attributed mostly to the
backreaction of the accelerated particles on the field through
injection. This effect is certainly present but as we showed here by
using only a hydrodynamical approach, the smoothening is in fact
mainly due to the reaction of the magnetic pressure on the background
plasma.

The smoothening of the precursor also results in two important
effects: 1) spectra of accelerated particles closer to power laws,
though the concavity which is peculiar of NL DSA remains evident. 2)
The maximum momentum of accelerated particles for given Mach number is
predicted to be somewhat larger (see \cite{bac07} for a detailed
discussion). Both these effects will be discussed in detail in a
forthcoming paper.  

\acknowledgments{The authors are grateful to the referee, Don Ellison, 
for his precious comments. This work was partially supported by 
PRIN-2006, by ASI through contract ASI-INAF I/088/06/0 and (for PB) by
the US DOE and by NASA grant NAG5-10842. Fermilab is operated by Fermi
Research Alliance, LLC under Contract No. DE-AC02-07CH11359 with the
United States DOE.}


\begin{thebibliography}{99}

\bibitem[\protect\citeauthoryear{Amato \& Blasi}{2005}]{amato1}
Amato, E. and Blasi, P., 2005, MNRAS Lett., 364, 76

\bibitem[\protect\citeauthoryear{Amato \& Blasi}{2006}]{amato2}
Amato, E. and Blasi, P., 2006, MNRAS, 371, 1251

\bibitem[\protect\citeauthoryear{Ballet}{2006}]{ballet}
Ballet, J., 2006, Adv. Sp. Res., 37, 1902

\bibitem[\protect\citeauthoryear{Bell}{1978a}]{bell78}
Bell, A.R., 1978a, MNRAS, 182, 147

\bibitem[\protect\citeauthoryear{Bell}{2004}]{bell2004}
Bell, A.R., 2004, MNRAS 353, 550

\bibitem[\protect\citeauthoryear{Bell \& Lucek}{2001}]{bl2001}
Bell, A.R. and Lucek S.G., 2001, MNRAS 321, 438

\bibitem[\protect\citeauthoryear{Berezhko \& Ellison}{1999}]{be99}
Berezhko E.G. and Ellison D.C., 1999, ApJ, 526, 385

\bibitem[\protect\citeauthoryear{Berezhko \& V\"{o}lk}{1997}]{bv97}
Berezhko E.G. and V\"{o}lk, H.J., 1997, APh, 7, 183

\bibitem[\protect\citeauthoryear{Blasi, Amato \& Caprioli}{2007}]{bac07}
Blasi, P., Amato, E. and Caprioli, D., 2007, MNRAS, 375, 1471B

\bibitem[\protect\citeauthoryear{Drury}{1983}]{druryrev}
Drury, L.O'C, 1983, Rep. Prog. Phys., 46, 973

\bibitem[\protect\citeauthoryear{Drury \& Falle}{1986}]{df86}
Drury, L.O'C and Falle, S.A.E.G., 1986, MNRAS, 223, 353D

\bibitem[\protect\citeauthoryear{Drury \& V\"{o}lk}{1981}]{dr_v81}
Drury, L.O'C and V\"{o}lk, H.J., 1981, ApJ, 248, 344

\bibitem[\protect\citeauthoryear{Giacalone \& Jokipii}{2007}]{gj07}
Giacalone J. and Jokipii J.R., 2007, ApJ, 663, L41

\bibitem[\protect\citeauthoryear{Jones \& Ellison}{1991}]{je91}
Jones, F.C., and Ellison, D.C., 1991, Sp. Sci. Rev., 58, 259

\bibitem[\protect\citeauthoryear{Malkov \& Drury} {2001}]{maldrury}
Malkov, M.A., Drury, L.O'C, 2001, Rep. Progr. Phys., 64, 429 

\bibitem[\protect\citeauthoryear{McKenzie \& V\"{o}lk}{1982}]{mck-v82}
McKenzie, J.F., and V\"{o}lk, H.J., A\&A, 116, 191

\bibitem[\protect\citeauthoryear{Parizot et al.}{2006}]{P+06} 
Parizot, E. et al., A\&A, 453, 387{}

\bibitem[\protect\citeauthoryear{Vainio \& Schlickeiser}{1999}]{vs99}
Vainio C., and Schlickeiser R., 1999, A\&A, 343, 303

\bibitem[\protect\citeauthoryear{Vladimirov, Ellison \&
    Bykov}{2006}]{elli06} 
Vladimirov, A., Ellison, D.C. and Bykov, A., 2006, ApJ, 652, 1246

\bibitem[\protect\citeauthoryear{V\"{o}lk, Berezhko \&
Ksenofontov}{2005}]{V+05}   V\"{o}lk H. J. , E. G. Berezhko and
L. T. Ksenofontov, A\&A, 433, 229 
    
\bibitem[\protect\citeauthoryear{V\"{o}lk \& McKenzie}{1981}]{v-mck81}
V\"{o}lk, H.J., and McKenzie, J.F., 1981, ICRC, 9, 246V

\bibitem[\protect\citeauthoryear{Warren et al}{2005}]{warren} 
Warren, J.S. et al., 2005, ApJ, 634, 376

\end{thebibliography}
\end{document}